\documentclass[12pt]{article} 
\usepackage{fullpage}

\usepackage{graphicx} 
\usepackage{amsmath, amsfonts, amsthm, amssymb} 
\usepackage{bbold} 
\usepackage{subfigure}
\usepackage{verbatim} 
\usepackage{hyperref}   
\usepackage{graphicx} 

\newcommand{\braket}[2]{\left\langle{#1}\vert{#2}\right\rangle} 
\newcommand{\ketbra}[2]{\left\vert{#1}\rangle\langle{#2}\right\vert} 
\newcommand{\selfbraket}[1]{\braket{#1}{#1}} 
\newcommand{\selfketbra}[1]{\ketbra{#1}{#1}} 

\DeclareMathOperator{\invsqrt2}{\frac{1}{\sqrt2}} 
\DeclareMathOperator{\id}{\mathbb{1}} 

\newcommand{\frob}[1]{\left\|{#1}\right\|_F}
\newcommand{\tr}[1]{\mathrm{Tr}\!\left[{#1}\right]} 
\DeclareMathOperator{\initstate}{\sum_{i=0}^3 A_i \otimes \sigma_i} 
\DeclareMathOperator{\finalexpander}{\mathcal{E}} 
\DeclareMathOperator{\depolarizer}{\mathcal{D}} 

\newcommand{\ceil}[1]{{\left\lceil{#1}\right\rceil}}

\newcommand{\set}[1]{{\lbrace{#1}\rbrace}}

\newcommand{\<}{\langle}      
\renewcommand{\>}{\rangle}

\newcommand{\N}{\mathbb{N}}

\newcommand{\C}{\mathbb{C}}

\newcommand{\calA} {\mathcal{A}}

\newcommand{\calH} {\mathcal{H}}

\newcommand{\calW} {\mathcal{W}}

\newcommand{\trP}{\mathrm{Tr}} 

\theoremstyle{plain}
\newtheorem{definition}{Definition}[section]
\newtheorem{theorem}[definition]{Theorem}
\newtheorem{proposition}[definition]{Proposition}

\theoremstyle{definition}
\newtheorem{remark}[definition]{Remark}

\input{Qcircuit}

\title{Testing quantum expanders is co-QMA-complete}

\date{October 2, 2012}

\author{Adam D. Bookatz\thanks{Center for Theoretical Physics, Massachusetts Institute of Technology, Cambridge, MA, USA; \texttt{bookatz@mit.edu}} \quad 
Stephen P. Jordan\thanks{National Institute of Standards and Technology, Gaitherburg, MD, USA; \texttt{stephen.jordan@nist.gov}} \quad 
Yi-Kai Liu\thanks{National Institute of Standards and Technology, Gaithersburg, MD, USA; \texttt{yi-kai.liu@nist.gov}} \quad 
Pawel Wocjan\thanks{Mathematics Department \& Center for Theoretical Physics, Massachusetts Institute of Technology, Cambridge, MA, USA; on sabbatical leave from Department of Electrical Engineering and Computer Science, University of Central Florida, Orlando, FL, USA; \texttt{wocjan@eecs.ucf.edu}}}

\begin{document}

\maketitle
  
\abstract{
A quantum expander is a unital quantum channel that is rapidly mixing, has only a few Kraus operators, and can be implemented efficiently on a quantum computer. We consider the problem of estimating the mixing time (i.e., the spectral gap) of a quantum expander. We show that this problem is co-QMA-complete. This has applications to testing randomized constructions of quantum expanders, and studying thermalization of open quantum systems.
%
}


%
%
\section{Introduction}

A quantum expander is a unital quantum channel that is rapidly mixing. This means that, with repeated applications of the channel, every quantum state is rapidly contracted to the maximally mixed state, which is the unique fixed point. 
In addition, a quantum expander has only a small number of Kraus operators, each of which is described by an efficient quantum circuit. 
Quantum expanders are quantum analogues of expander graphs, which play a prominent role in computer
science and discrete mathematics \cite{HLW06}.  The idea of quantum
expanders was introduced in \cite{Hastings07,BT07}.  
Since then, several explicit constructions of quantum expanders have been discovered, and quantum expanders have found various applications in quantum information theory, such as constructing quantum states with unusual entanglement properties, and simulating thermalization in quantum systems \cite{BST07,Hastings07b,GE08,Harrow08,HH09,BHH12}.

Here we study the problem of estimating the mixing rate of a quantum expander.  Given a quantum channel $\Phi$ of the above form (a small number of Kraus operators, specified by quantum circuits), this problem is to estimate the spectral gap of $\Phi$.  This problem arises in connection with randomized constructions of quantum expanders \cite{BHH12}, where with high probability one obtains a good expander, but it is not obvious how to test that a particular instance of the construction is in fact good.  In addition, this problem can be viewed as a special case of a more general question: given an open quantum system, determine whether it thermalizes, and on what time scale.  (The behavior of a quantum expander is roughly equivalent to that of a quantum system with a particular weak coupling to a bath of harmonic oscillators.)

Formally, we define the ``quantum non-expander problem'' (which is the complement of the above problem), and we give evidence that this problem is computationally intractable:  we prove that it is QMA-complete. 
Here QMA (Quantum Merlin-Arthur) is a complexity class that is a quantum analogue of NP (Nondeterministic Polynomial Time) \cite{Knill96, KSV02, Watrous00}.  Proving that a problem is QMA-complete implies that it is equivalent (up to polynomial-time reductions) to all other QMA-complete problems \cite{KSV02, JWB05, Liu06, LCF07, BS07, KKR06, AGIK09, shortNIC}, a survey of which can be found in \cite{Bookatz12}.  In particular, this implies that the problem cannot be solved in polynomial time (unless QMA = BQP).  Furthermore, this implies that our original problem, the ``quantum expander problem,'' cannot be in QMA (unless QMA = coQMA).  In other words, when a channel $\Phi$ is \emph{not} a quantum expander, there is an efficiently-verifiable quantum proof of that fact; but when $\Phi$ \emph{is} a quantum expander, there is no way of giving an efficiently-verifiable quantum proof.

%
%
  
\section{Preliminaries}\label{sec:defNEP}

\subsection{The quantum non-expander problem}
  
We use the definition of explicit quantum expanders due to Ben-Aroya, Schwartz, and Ta-Shma \cite{BST07}.
For an $N$-dimensional Hilbert space $\calH$, let $L(\calH)$ denote the space of linear operators from $\calH$ to itself.  
A superoperator $\Phi : L(\calH) \rightarrow L(\calH)$ is admissible if it is a completely positive and trace-preserving map.  An admissible superoperator is unital
if $\Phi(\tilde{I})=\tilde{I}$, where $\tilde{I}=\frac{I}{N}$ is the maximally mixed state on $\calH$ (where $I$ is the identity operator on $\calH$).  A unital superoperator is $D$-regular if $\Phi=\frac{1}{D} \sum_d \Phi_d$, and for $d=1,\ldots,D$, $\Phi_d(X)=U_d X U_d^\dagger$ where the $U_d$ are unitary transformations on $\calH$. The unitaries $U_d$ are called the operation elements (or Kraus operators) of $\Phi$, and $D$ is called the degree of $\Phi$.  A $D$-regular superoperator is explicit
if each of its operation elements can be implemented by a quantum circuit of size $\mathrm{polylog}(N)$, where $N$ is the dimension of $\calH$.
  
%
%
  
\begin{definition}[Quantum expander] \label{def:expander}
A $D$-regular superoperator $\Phi : L(\calH) \rightarrow L(\calH)$ is a $\kappa$-contractive expander if
for all $A\in L(\calH)$ that are orthogonal to $\tilde{I}$ with respect to the Hilbert-Schmidt inner product, that is, $\trP(A \tilde{I})=0$, it holds that
\begin{equation}\label{eq:expander_defn}
\|\Phi(A)\|_F \le \kappa \|A\|_F.
\end{equation}
Here the Frobenius norm is given by $\|A\|_F=\sqrt{\sum_{i,j} |a_{ij}|^2}$, where $a_{ij}$ are the entries of the matrix $A$.  
The quantity $1-\kappa$ is called the spectral gap of $\Phi$.
\end{definition}

\begin{remark} \label{remark:motivation}
The motivation for this definition can easily be seen from the following argument. A good quantum expander $\Phi$ rapidly sends any density matrix $\rho$ to the maximally mixed state $\tilde{I}$. Because $\tr{\rho}=\tr{\tilde{I}}=1$ we can always write $\rho = \tilde{I} + A$ where $\tr{A}=0$. The requirement of Eq.~(\ref{eq:expander_defn}) therefore formalizes the idea of $\Phi$ bringing $\rho$ towards $\tilde{I}$ by rapidly killing off the $A$ term. In this context Eq.~(\ref{eq:expander_defn}) is equivalent to demanding that $\frob{\Phi(\rho)-\tilde{I}} \leqslant \kappa \frob{\rho-\tilde{I}}$, which clearly encapsulates the idea of  $\Phi$ rapidly sending density matrices towards the maximally mixed state. Note that in this argument $A=\rho-\tilde{I}$ is Hermitian; however, it can be shown that if Eq.~(\ref{eq:expander_defn}) applies for traceless Hermitian matrices, it also applies for traceless matrices in general, thus justifying Definition~\ref{def:expander}.
\end{remark}

We consider the problem of estimating the mixing time of a quantum expander.  Formally, we study the following decision problem:

%
%

\begin{definition}[Quantum non-expander problem] \label{def:nonexpander_problem}
Fix some encoding such that each string $x \in \set{0,1}^*$ specifies the following: an explicit $D$-regular superoperator $\Phi : (\C^2)^{\otimes m} \rightarrow (\C^2)^{\otimes m}$, with operation elements $U_1,\ldots,U_D$, and two parameters $\alpha > \beta$. 

We will consider instances which satisfy the following promises\footnote{Here $|x|$ denotes the length of the string $x$.}: $m$ and $D$ are upper-bounded by (fixed) polynomials in $|x|$; the parameters $\alpha$ and $\beta$ are polynomially separated, i.e., they satisfy $\alpha - \beta \ge \frac{1}{q(|x|)}$ for some (fixed) polynomial $q$; and the operation elements $U_1,\ldots,U_D$ are given as quantum circuits of size at most $r(|x|)$ for some (fixed) polynomial $r$.

The ``quantum non-expander'' problem is the task
of deciding which of the following is correct, given the promise that
exactly one of them is correct:
\begin{itemize}
\item $\Phi$  is not an $\alpha$-contractive expander (YES case)
\item $\Phi$ is a $\beta$-contractive expander (NO case)
\end{itemize}
\end{definition}  

\subsection{Thermalization of open quantum systems}

To motivate the ``quantum non-expander'' problem, we now describe a connection between that problem and the study of thermalization in open quantum systems.  We show an example of a quantum system coupled to a bath, where the system thermalizes, and the relaxation time is determined by the spectral gap of a certain quantum expander.  

Let the system consist of $m$ qubits, and fix some unitary transformations $U_\alpha$ (for $\alpha=1,\ldots,D$) which act on $(\C^2)^{\otimes m}$.  Let the bath consist of a large number of harmonic oscillators, with annihilation operators $b_{\alpha k}$ (for $\alpha=1,\ldots,D$ and $k\in\Omega$, where $\Omega$ is some large set).  Let the total Hamiltonian be 
\begin{equation} \label{eqn-ham1}
H = H_S + \varepsilon H_I + H_B, 
\end{equation}
where the system Hamiltonian is $H_S=0$, the bath Hamiltonian is 
\begin{equation} \label{eqn-ham2}
H_B = \sum_\alpha \sum_k \omega_k b_{\alpha k}^\dagger b_{\alpha k}, 
\end{equation}
and the interaction Hamiltonian is 
\begin{equation} \label{eqn-ham3}
H_I = \sum_\alpha (U_\alpha \otimes f_\alpha) + (U_\alpha^\dagger \otimes f_\alpha^\dagger), 
\end{equation}
where the operators $f_\alpha$ are defined by $f_\alpha = \frac{1}{\sqrt{|\Omega|}} \sum_k b_{\alpha k}$.

In the weak-coupling limit ($\varepsilon \rightarrow 0$), the time evolution of the system is described by a master equation \cite{BP-book}.  Suppose the bath is in a thermal state, $\rho_B = (1/Z_B) \exp(-H_B/T)$.  Then the master equation takes the following form:
\begin{equation}\label{eqn-master}
\frac{d}{dt} \rho_S(t)
 = R_0 \sum_\alpha \Bigl( U_\alpha \rho_S(t) U_\alpha^\dagger - \rho_S(t) \Bigr)
 + R_1 \sum_\alpha \Bigl( U_\alpha^\dagger \rho_S(t) U_\alpha - \rho_S(t) \Bigr), 
\end{equation}
where $\rho_S(t)$ is the state of the system at time $t$, and $R_0$ and $R_1$ are positive real numbers.  Equation (\ref{eqn-master}) has two special features:  there is no contribution from a ``Lamb shift'' Hamiltonian, and the dissipator is in diagonal form with Lindblad operators which are unitary.  (See Appendix \ref{apx-thermal} for the derivation of this equation.)

Now define the quantum channel 
\[
\Phi(\rho)
 = \frac{R_0}{(R_0+R_1)D} \sum_\alpha U_\alpha \rho U_\alpha^\dagger
 + \frac{R_1}{(R_0+R_1)D} \sum_\alpha U_\alpha^\dagger \rho U_\alpha. 
\]
This channel $\Phi$ is a (non-uniform) mixture of unitary operations.  In the special case where the set of unitaries $\set{U_\alpha \;|\; \alpha=1,\ldots,D}$ is closed with respect to the adjoint operation (i.e., for every $1 \leq \alpha \leq D$, there exists some $1 \leq \beta \leq D$ such that $U_\alpha = U_\beta^\dagger$), the channel $\Phi$ can be written as 
\[
\Phi(\rho) = \frac{1}{D} \sum_\alpha U_\alpha^\dagger \rho U_\alpha,
\]
hence $\Phi$ is a $D$-regular superoperator, as described in the definition of a quantum expander.

The master equation can now be rewritten in terms of $\Phi$:
\[
\frac{d}{dt} \rho_S(t) = (R_0+R_1)D \cdot \bigl(\Phi-\mathcal{I}\bigr) (\rho_S(t)), 
\]
where $\mathcal{I}$ denotes the identity channel.  We can solve for $\rho_S(t)$:
\[
\rho_S(t) = \exp\Bigl( t \cdot (R_0+R_1)D \cdot \bigl(\Phi-\mathcal{I}\bigr) \Bigr) (\rho_S(0)).
\]
Thus the system converges to the maximally mixed state as $t \rightarrow \infty$, and the rate of convergence depends on the spectral gap of $\Phi$.  More precisely, write $\rho_S(t) = \tilde{I} + A(t)$ where $A(t)$ is traceless.  Then it can be verified that
\[
\| A(t) \|_F \le \exp\big(-t \cdot (R_0+R_1)D (1-\kappa)\big) \|A(0)\|_F.
\]

\subsection{Quantum Merlin-Arthur}

We will show that the quantum non-expander problem is QMA-complete, i.e., it is contained in QMA, and every problem in QMA can be reduced to it in polynomial time.

The complexity class QMA consists of decision problems such that YES
instances have concise quantum proofs. The name QMA stands for Quantum
Merlin-Arthur, which is motivated by the following protocol. Given a
problem instance $x$ (\emph{i.e.} a string of $|x|$ bits), and a
language $L \in QMA$, a computationally unbounded but untrustworthy
prover, Merlin, submits a quantum state of $\mathrm{poly}(|x|)$ qubits
as a purported proof that $x \in L$. A verifier, Arthur, who can
perform polynomial size quantum computations, then processes this
proof and either accepts or rejects it. If $x \in L$ then there exists
some polynomial size quantum state causing Arthur to accept with high
probability, but if $x \notin L$ then Arthur will reject all states
with high probability. QMA is a quantum analogue of MA, which is the
probabilistic analogue of NP. 


\begin{definition}[QMA$(a,b)$]
A language $L$ is in QMA$(a,b)$ if for each $x\in \{0,1\}^*$ one can efficiently
generate a quantum circuit $V$ with the following properties:
\begin{itemize}
\item $V$ acts on the Hilbert space $\calW \otimes \calA$ where 
\[
\calW = (\C^2)^{\otimes n_w}, \quad \calA = (\C^2)^{\otimes n_a},
\]
and the functions $n_w, n_a : \N \rightarrow \N$ grow at most polynomially in $|x|$
\item $V$ consists of $s(|x|)$ elementary gates where the function $s : \N \rightarrow \N$ grows at most polynomially in $|x|$
\item if $x\in L$ (YES case) then there exists a witness state $\ket{\psi}\in \calW$ such that 
\begin{equation}\label{eq:QMA_yes}
\left\Vert P V \ket{\psi}\ket{\mathbf{0}}\right\Vert^2 \ge a
\end{equation}
\item
if $x\notin L$ (NO case) then for all states $\ket{\psi}\in \calW$ we have that 
\begin{equation}\label{eq:QMA_no}
\left\Vert P V \ket{\psi}\ket{\mathbf{0}}\right\Vert^2 \le b
\end{equation}
\end{itemize}
Here $\calW$ and $\calA$ are the witness and ancilla registers,
respectively, and $P = \selfketbra{1}\otimes \id$ projects onto the
subspace of the first qubit of $\calW\otimes\calA$ being in the state
$|1\>$.  The state $\ket{\mathbf{0}}=\ket{00\ldots 0}$ is the
all-zeros state on $\calA$. 
\end{definition}
Observe that $V,\calW, \calA, n_a, n_w$ and $P$ depend on $x$;
however, to avoid unnecessarily complicated notation, we do not
indicate this explicitly.

\begin{remark} \label{remark:amplification}
It is conventional to define $\mathrm{QMA} = \mathrm{QMA}(2/3,
1/3)$. However, the complexity class $\mathrm{QMA}(a,b)$ is highly
insensitive to the particular values of $a$ and $b$. In fact, even if
$a$ and $b$ are functions of the problem size $n$, it remains true
that $\mathrm{QMA}(a(n),b(n)) = \mathrm{QMA}$ provided $a(n) - b(n)
\geq \frac{1}{p(n)}$ for some polynomial $p$. It is always possible to
achieve that $a=1-\varepsilon$ and $b=\varepsilon$ by increasing the
size of the circuit by a factor $\mathrm{polylog}(1/\varepsilon)$ and
increasing $n_a$ by $\mathrm{polylog}(1/\varepsilon)$ qubits, with no
change in $n_w$ \cite{MW05,amplificationQMA}. 
\end{remark}

%
%

\section{Quantum non-expander is in QMA}\label{sec:inQMA}

We now show that the problem defined in Definition~\ref{def:nonexpander_problem} is in QMA.  We first consider the YES case.  In this case, Merlin has to convince Arthur that there exists a traceless matrix $A$ such that 
\begin{equation}\label{eq:badContraction}
\| \Phi(A)\|_F > \alpha \|A\|_F.
\end{equation}
We may assume w.l.o.g. that $\|A\|_F=1$.  Clearly, Merlin cannot directly send the matrix $A$ because it is an exponentially large matrix.  Instead, he can send the quantum certificate
\[
|\psi_A\> = \sum_{i,j=1}^N a_{ij} |i\> \otimes |j\>
\]
encoding the matrix $A$.
We show that $|\psi_A\>$ can serve as a witness making it possible to convince Arthur that the inequality in Eq.~(\ref{eq:badContraction}) holds.

Arthur's verification protocol makes use of the following facts:
\[
\|A\|_F^2 = \<\psi_A|\psi_A\>,
\]
\[
\tr{A} = {\sqrt{N}} \<\varphi|\psi_A\>,
\]
where $|\varphi\> = \frac{1}{\sqrt{N}} \sum_{i=1}^N |i\> \otimes |i\>$, and
\[
\|\Phi(A)\|_F^2 = \< \psi_A|W^\dagger W |\psi_A\>,
\]
where 
\[
W = \frac{1}{D} \sum_{d=1}^D U_d \otimes \overline{U}_d
\]
and $\overline{U}_d$ denotes the complex conjugate of $U_d$.  

First, to check whether $\tr{A}=0$, Arthur verifies that $|\psi_A\>$ is orthogonal to $\ket{\varphi}$.  Second, to estimate the contractive factor, Arthur estimates the expectation value $\<\psi_A|W^\dagger W|\psi_A\>$ of $W^\dagger W$.  For $d,e=1,\ldots,D$, define the unitaries
\[
V_{d,e} = (U_d^\dagger \otimes U_d^T) (U_e \otimes \overline{U}_e).
\]
Note that $V_{d,e}=V_{e,d}^\dagger$ and $V_{d,d}=\id$.  The expectation value can be expressed as
\[
\<\psi_A|W^\dagger W|\psi_A\> = \frac{1}{D^2} \sum_{d,e} \<\psi_A|V_{d,e}|\psi_A\> =
\frac{1}{D} + \frac{2}{D^2} \sum_{d<e} \mathrm{Re} \<\psi_A|V_{d,e}|\psi_A\>.
\]
\begin{figure}
	\begin{center}
		\includegraphics[height=0.8in]{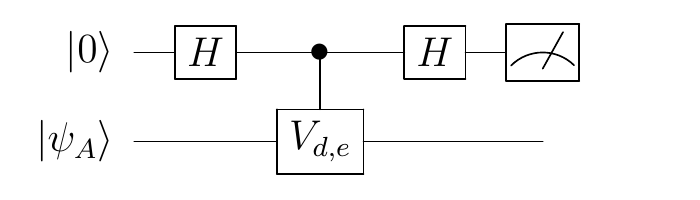}
		\caption{Hadamard test for $V_{d,e}$}
		\label{fig:hadamard_test}
	\end{center}
\end{figure}
Arthur can estimate the values $\mathrm{Re}\<\psi_A|V_{d,e}|\psi_A\>$ using the Hadamard test [shown in Fig.~(\ref{fig:hadamard_test})] since it will output $0$ with probability $\Pr(0)=\frac{1}{2}(1+\mathrm{Re}\<\psi_A|V_{d,e}|\psi_A\>)$. From this Arthur can calculate $\< \psi_A|W^\dagger W |\psi_A\> = \|\Phi(A)\|_F^2$ and ensure it exceeds $\alpha^2$.

Now consider the NO case. In this case, Arthur's first measurement projects the state $|\psi_A\>$ onto the subspace orthogonal to $|\varphi\>$; and by definition, all states $|\psi_A\>$ in that subspace must satisfy
\[
\<\psi_A|W^\dagger W|\psi_A\> = \|\Phi(A)\|_F^2 \leqslant \beta^2.
\]
This shows that Merlin cannot cheat, that is make Arthur believe that there exists a quantum state with contraction greater or equal to $\alpha$, provided that Arthur estimates the expected value sufficiently well and with sufficiently high probability of confidence.

As in the original definition of QMA in \cite{KSV02}, we may assume that Arthur has multiple copies of the quantum certificate $|\psi\>$ so that we can estimate the expected value sufficiently well.  
Using the powerful technique of in-place amplification \cite{MW05}, we can transform a quantum circuit requiring $|\psi\>^{\otimes k}$ into one that requires only a single copy of $|\psi\>$.
 
%
%
 
\section{Some technical tools}\label{facts}

\subsection{The Frobenius norm}

In the proof that quantum non-expander is QMA-hard we will frequently make use
of the Frobenius norm; we therefore present some useful facts about
this norm here. If $B$ is a
matrix with entries $b_{ij}$, then the Frobenius norm is defined as
\begin{equation}
\label{fact:frob_defn}
\frob{B} = \sqrt{\tr{B^\dag B}} = \sqrt{ \sum_{ij} |b_{ij}|^2}.
\end{equation}
We have the following identities: $\frob{A\otimes B} =
\frob{A}\frob{B}$, $\tr{A\otimes B}=\tr{A}\tr{B}$, and of course
$\tr{A+B}=\tr{A}+\tr{B}$. If $\ket{\psi}$ and $\ket{\phi}$ are pure
states then
\begin{equation}
\label{fact:norms}
\frob{\Big.\ketbra{\psi}{\phi}\Big.} = \sqrt{\selfbraket{\psi}\selfbraket{\phi}} = \Big\| \ket{\psi} \Big\|\Big\| \ket{\phi} \Big\|.
\end{equation}
Note that $\frob{\big.\selfketbra{0}\big.} = \frob{\big.\selfketbra{1}\big.} = 1$.

In this paper we denote the Pauli matrices on one qubit by $\sigma_i$,
with $\sigma_0=\mathbb{1}$, $\sigma_1 = \sigma_x$, $\sigma_2 =
\sigma_y$, and $\sigma_3 = \sigma_z$. Consider any traceless matrix
$A$ that acts on some space $\mathbb{C}^d\otimes \mathbb{C}^2$, where
we will refer to the second subspace (i.e. single-qubit subspace) as
the \textit{indicator qubit register}. Because the Pauli matrices
$\sigma_i$ form a basis for the matrices acting on the indicator qubit
register, we can decompose $A$ as $\sum_{i=0}^3 A_i \otimes \sigma_i$,
where $A_i$ are matrices on the combined multiqubit subspace (the
witness and ancilla registers that we will see later). Because
$\sigma_i$ are traceless for $i=1,2,3$, the traceless condition on $A$
therefore becomes $\tr{A_0}=0$. Moreover, because the Pauli matrices are
orthogonal with respect to the trace inner product and all satisfy
$\frob{\sigma_i}^2=2$, we have $\frob{\sum_i A_i \otimes \sigma_i}^2 =
\sum_i \frob{A_i \otimes \sigma_i}^2 = 2\sum_i \frob{A_i}^2$, giving
the inequality 
\begin{equation}
\label{fact:sum_greater_than_part}
\frob{\initstate} \geqslant \sqrt2\frob{A_0}.
\end{equation}

A quantum operation $G$ is called a pinching operator if $G(B) =
\sum_P PBP$ where $P$ are non-overlapping projectors with $\sum_P P =
\mathbb{1}$. Pinching operators are trace preserving,
\begin{equation}
\label{fact:pinch_trace}
\tr{\sum_P PBP} = \tr{B},
\end{equation}
and moreover, (by the pinching inequality) cannot increase Frobenius
norm:
\begin{equation}
\label{fact:pinch_norm}
\frob{\sum_P PBP} \leqslant \frob{B}.
\end{equation}
It should be noted that a quantum expander $\finalexpander$ is also
norm-non-increasing, 
\begin{equation}
\label{fact:expander_decreases}
\frob{\finalexpander(B)} \leqslant \frob{B},
\end{equation}
and similarly for any projector $P$,
\begin{equation}
\label{fact:projector_decreases}
\frob{PBP} \leqslant \frob{B}.
\end{equation}

\subsection{Controlled expanders}

The remainder of our paper will make repeated use of controlled expanders, which we introduce here. If $U$ is a unitary gate, we use the notation $\Lambda U$ to indicate a controlled-$U$ operation.
  
\begin{definition}[Controlled expander]
Let $\mathcal{F}$ be a quantum expander with operation elements $\lbrace U_i : i=1\ldots m\rbrace$ so that $\mathcal{F}(B) = \frac{1}{m}\sum_{i=1}^m U_i B U_i^\dag$. 
The controlled expander $\Lambda\mathcal{F}$ is defined to be the $m$-regular superoperator whose operation elements are the controlled unitaries $\lbrace \Lambda U_i : i=1\ldots m\rbrace$. 
\end{definition}

More explicitly, consider two registers, a control register and a target register, and suppose that an expander $\mathcal{F}$ acts on the target register as $\mathcal{F}(B) = \frac{1}{m}\sum_{i=1}^m U_i B U_i^\dag$. Decompose the control register into two orthogonal subspaces, and let $Q$ and $P$ be projectors onto these two subspaces (so $Q+P=\id$ and $PQ=QP=0$). Suppose that the controlled operations $\Lambda U_i$ are to be applied when the control register is in the subspace corresponding to $P$; thus $\Lambda U_i = P\otimes U_i + Q\otimes \mathbb{1}$. Consider a matrix $A \otimes B$, where $A$ and $B$ act on the control and target registers, respectively. Then the controlled expander $\Lambda\mathcal{F}$, with operation elements $\Lambda U_i$, acts on $A \otimes B$ as
\begin{eqnarray}
\label{eq:controlled_expander_long}
\Lambda\mathcal{F}\left(A \otimes B \right)
& = & 
\frac{1}{m} \sum_{i=1}^m \Big[(\Lambda U_i) (A\otimes B) (\Lambda U_i^\dag) \Big]\notag\\
& = & 
\frac{1}{m} \sum_{i=1}^m \Big[(P\otimes U_i + Q\otimes \mathbb{1}) (A\otimes B) (P\otimes U_i^\dag + Q\otimes \mathbb{1}) \Big]\notag\\
& = & 
\frac{1}{m} \sum_{i=1}^m \Big[P A P \otimes U_i B U_i^\dag + P A Q \otimes U_i B + Q A P \otimes B U_i^\dag + Q A Q \otimes B \Big] \notag\\
& = & 
P A P \otimes \frac{1}{m} \sum_i (U_i B U_i^\dag)+ P A Q \otimes \left(\frac{1}{m} \sum_i U_i\right)B \\
 && ~~~~~~~~+ Q A P \otimes B \left(\frac{1}{m} \sum_i U_i^\dag\right) + Q A Q \otimes B .\notag
\end{eqnarray}

Note that if we impose on $\mathcal{F}$ the requirement that 
\begin{equation}
\label{eq:controlled_condition}
\sum_i U_i = 0
\end{equation}
then we obtain
\begin{equation}
\label{eq:controlled_expander_short}
\Lambda\mathcal{F}\left(A \otimes B \right) 
= P A P \otimes \mathcal{F}(B) + Q A Q \otimes B
\end{equation} 
which is how we would naturally desire a controlled expander to
act. Unfortunately, unlike Eq.~(\ref{eq:controlled_expander_short}),
Eq.~(\ref{eq:controlled_expander_long}) has additional crossterms
whose elimination would greatly simplify our future analysis.

We will, however, freely assume that Eq.~(\ref{eq:controlled_condition}) is satisfied, justified by the following observation.
If necessary, we may always increase the set of operation elements of $\mathcal{F}$ from $\lbrace U_i
: i=1\ldots m\rbrace$ to $\lbrace U_i : i=1\dots m\rbrace\cup\lbrace
-U_i : i=1\ldots m\rbrace$. Such a change has no effect on the original expander $\mathcal{F}$; the expander $\mathcal{F}(B) = \frac{1}{m} \sum(U_i ~B~
U_i^\dag$) is invariant under $U_i \leftrightarrow -U_i$, even though the
controlled expander $\Lambda\mathcal{F}(B) = \frac{1}{m} \sum(\Lambda
U_i ~B~ \Lambda U_i^\dag$) is not necessarily invariant under $U_i
\leftrightarrow -U_i$.
Thus, with only a factor of two overhead in the number of unitaries, we may satisfy the condition of Eq.~(\ref{eq:controlled_condition}), thereby eliminating the undesired crossterms; as such, Eq.~(\ref{eq:controlled_expander_short}) may effectively be taken as the definition of a controlled expander.


A concrete example of a controlled expander -- and one of particular
importance in this paper -- is the \textit{controlled complete
  depolarizer}. Throughout this paper we use $\depolarizer$ to denote
the complete depolarizing channel on a single qubit, which is normally
defined to apply a unitary from $\lbrace \id,X,Y,Z \rbrace$ with
uniform probability 1/4. To ensure that
Eq.~(\ref{eq:controlled_condition}) is satisfied, we therefore define
the effect of $\depolarizer$ on a matrix $\sigma$ to be
\[
\depolarizer(\sigma) = \frac{1}{8}\sum_W W\sigma W = \id\frac{\tr \sigma}{2}
\]
 where the sum is over $W\in \lbrace \id,X,Y,Z,-\id,-X,-Y,-Z \rbrace$. 
Consequently, the controlled complete depolarizer $\Lambda\depolarizer$ with a single qubit target and (possibly multiqubit) control projectors $P$ (indicating apply $\depolarizer$) and $Q$ (indicating do nothing) is the 8-regular superoperator with operation elements 
\[
\lbrace \Lambda(\id),\Lambda(X),\Lambda(Y),\Lambda(Z),\Lambda(-\id),\Lambda(-X),\Lambda(-Y),\Lambda(-Z) \rbrace
\] having the effect
\begin{equation}
\label{eq:controlled_depolarizer}
\Lambda\depolarizer(A\otimes\sigma) = P A P \otimes \mathbb{1}\frac{\tr{\sigma}}{2} + Q A Q \otimes \sigma .
\end{equation}

\begin{figure}
	\begin{center}
		\includegraphics[height=0.6in]{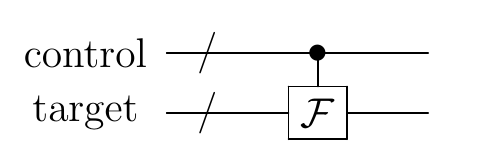}
		\caption{A controlled expander, $\Lambda\mathcal{F}$}
		\label{fig:controlled_expander}
	\end{center}
\end{figure}

Although controlled expanders are not actually quantum gates, we will nevertheless include them in circuit diagrams. If $\Lambda \mathcal{F}(B) = \frac{1}{m} \sum_i(\Lambda U_i ~B~ \Lambda U_i^\dag)$ then the circuit in Fig.~\ref{fig:controlled_expander} is to be interpreted as applying an element selected uniformly at random from the set $\lbrace \Lambda U_i \rbrace$ (or equivalently, as applying to the target register a unitary selected uniformly at random from the set $\lbrace U_i \rbrace$, but only if the control register is in the appropriate state.). As a final remark note that although a controlled expander is a unital map, it is not itself a good expander (firstly, because depending on the control qubit, the operator might not do anything at all, and secondly because even when the operator does act, it only expands on the subspace of the target, not the entire space). For example, note that $\selfketbra{0}\otimes\selfketbra{0}$ is not contracted at all by the controlled complete depolarizer $\Lambda\depolarizer$, thus indicating that $\Lambda\depolarizer$ is not a good expander.

%
%

\section{Quantum non-expander is QMA-hard}\label{sec:QMAhard}

\subsection{Outline of the proof}

Let $L$ be any language in QMA($\frac{2}{3},\frac{1}{3}$). We show
that the quantum non-expander problem is QMA-hard by reducing $L$ to a
quantum non-expander problem. Specifically, let $x$ be an $|x|$-bit problem
instance whose inclusion in $L$, or lack-thereof, we wish to
determine. Because $L\in $ QMA we have access to a verifier circuit
satisfying Eqs.~(\ref{eq:QMA_yes}) and~(\ref{eq:QMA_no}) acting on a
witness space of $n_w=\text{poly}(|x|)$ qubits and some ancilla
space. For reasons that will become apparent later, we now use QMA
amplification to give that $L\in \text{QMA}(a,b)$ for polynomially
separated $a$ and $b$ where 
\begin{equation*}
a>0.99  \quad\mbox{and}\quad b < (0.1)2^{-(n_w + 1)}. 
\end{equation*}
Note from Remark~\ref{remark:amplification} that this can be done without increasing the size of the witness space of the verifier. Let the resulting QMA$(a,b)$ verifier circuit be called $V$, which acts on the same witness space of $n_w=\text{poly}(|x|)$ qubits and some ancilla space of $n_a=\text{poly}(|x|)$ ancilla qubits. Merlin can provide Arthur a valid (with high probability) witness if and only if $x\in L$. 

\begin{figure}
\begin{center}
\includegraphics[height=1.7in]{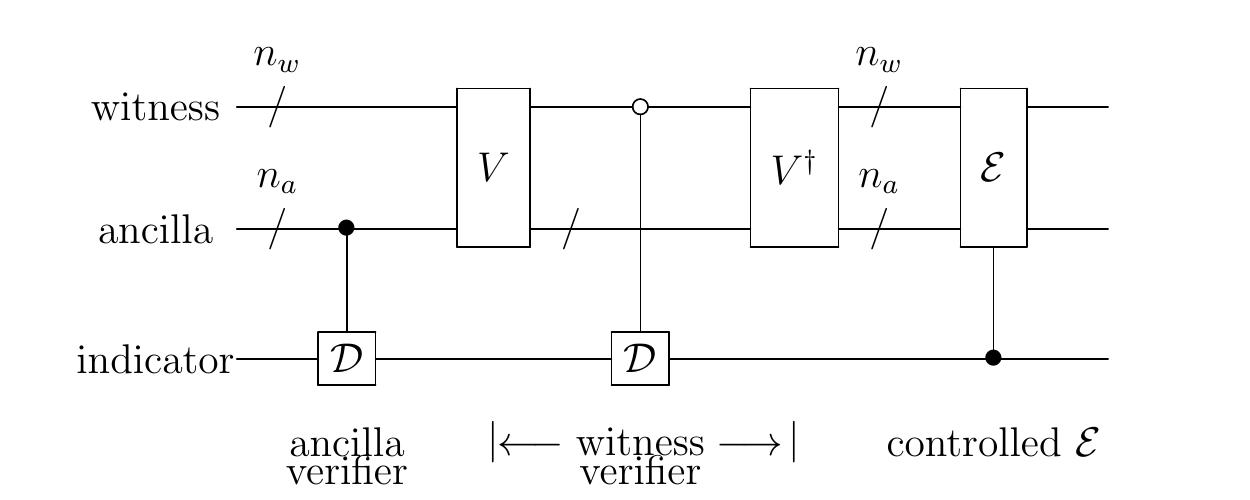}
\caption{The map $\Phi$ constructed from the verifier circuit $V$, the complete depolarizer $\depolarizer$, and the $\kappa_{\finalexpander}$-contractive expander $\finalexpander$. The first controlled depolarizer is applied only if the ancillae are not all zero and the second one only if the top output is zero.  The controlled $\finalexpander$-expander is applied only if the bottom qubit is one. Note that this figure is not a true circuit because $\depolarizer$ and $\finalexpander$ are quantum expanders, not unitary gates.}
\label{fig:whole_circuit}
\end{center}
\end{figure}

Let $\finalexpander$ be an explicit $\kappa_{\finalexpander}$-contracting expander of degree $D_{\finalexpander}$ acting on $n_w+n_a$ qubits, where $\kappa_{\finalexpander}< 0.1$ and $D_{\finalexpander}$ is constant (independent of $|x|$). 
 Such expanders are known to exist, as we outline in Appendix~\ref{sec:appendix} using Ref~\cite{BST10}.
Using $V$ and $\finalexpander$, we create a quantum expander $\Phi$ that is bad if $x\in L$ but good if $x\notin L$; indeed, we will present polynomially-separated (in fact, constant) $\alpha$ and $\beta$ such that $\Phi$ is a $\beta$-contracting expander if $x\notin L$ but is not an $\alpha$-contracting expander if $x\in L$. The circuit for $\Phi$ is shown in Fig.~\ref{fig:whole_circuit}, which we now describe in detail.


The map $\Phi$ acts on three registers, which from top to bottom are 
the witness register (of $n_w$ qubits), 
the ancilla register (of $n_a$ qubits), and 
an additional single-qubit register we call the \textit{indicator qubit} register.
The circuit is realized by composing the following three maps: 
\begin{enumerate}
\item the ancilla verifier
\item the witness verifier
\item the controlled $\finalexpander$.
\end{enumerate}

The basic idea is that if $x\in L$ then Merlin can provide a valid witness and properly initialized ancillae that will pass the verifiers and not be mixed by the final controlled expander (indicating that our quantum expander is bad); conversely, if $x\notin L$ then no matter what witness and ancilla qubits Merlin provides, 
the indicator qubit will be depolarized and consequently 
his state will be well-mixed by the final controlled expander (indicating our expander to be good).

We now provide a detailed description of the three different maps and their purposes.

\begin{enumerate}
\item 
The ancilla verifier is the first gate in Fig.~\ref{fig:whole_circuit}. It is the controlled expander $\Lambda_{anc} \depolarizer$, which applies the complete depolarizer $\depolarizer$ to the indicator qubit register only if any of the ancilla bits are 1 (i.e. if they are not all 0). More technically, it is 
\[
\Lambda_{anc} \depolarizer (B) = \frac{1}{8}\sum_W \Lambda_{anc} W ~B~ \Lambda_{anc} W^\dag
\]
(with $W\in \lbrace \id,X,Y,Z,-\id,-X,-Y,-Z \rbrace$), where $\Lambda_{anc} W$ is the gate shown in Fig.~\ref{fig:ancilla_verifier}. Note that $\Lambda_{anc} W$ requires a controlled-$W^\dag$ gate controlled by $n_a$ qubits, which can be implemented with ${n_a}^2$ gates using no extra work qubits~\cite{Elementary95}. (It is important that the implementation not require work qubits, because we demand that there are no internal ancillae; our expander must be an expander on the entire space, not just a subspace.)
Intuitively, if the ancilla qubits are not initialized to be all $0$'s, the verifier will depolarize the indicator qubit, whence the term ancilla verifier.

\begin{figure}
	\begin{center}
		\includegraphics[height=2in]{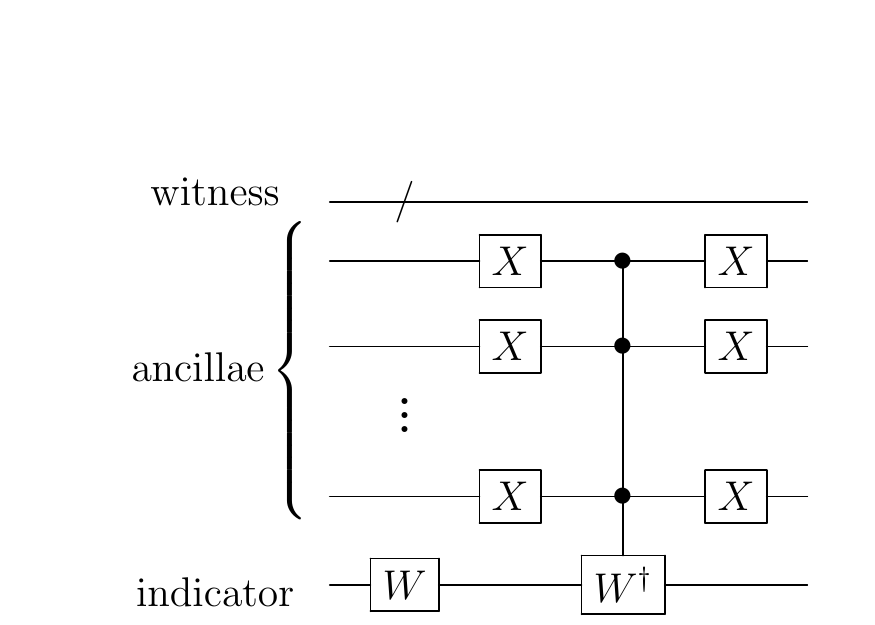}
\caption{The controlled expander verifying the ancillae. The unitary $W$ is selected from $\lbrace \id, X, Y, Z, -\id, -X, -Y, -Z\rbrace$ uniformly at random.}
\label{fig:ancilla_verifier}
	\end{center}
\end{figure}

\item 
The witness verifier consists of the next three operations in Fig.~\ref{fig:whole_circuit}. First, $V$ operates on the witness and ancilla registers, with its output on the top qubit (with $\ket{1}$ signifying that the witness is valid, $\ket{0}$ signifying that it is invalid); the lower multiqubit register on $n_w+n_a-1$ qubits contains the rest of $V$'s output (required by reversibility). A controlled-depolarizer then acts on the indicator qubit, conditioned upon the top qubit being $\ket{0}$ (i.e. failing the witness verification). The effects of $V$ are then uncomputed with $V^\dag$.
At this point, intuitively, the indicator qubit has been depolarized if and only if the input failed either the ancilla verifier or the witness verifier (or both). 

\item
Finally, the last gate, which is the controlled expander $\Lambda_{ind} \finalexpander$, acts, conditioned on whether the indicator qubit is $\ket{1}$. Intuitively, if the input was $\ket{\psi}\otimes\ket{\mathbf{0}}\otimes\ket{0}$, with the indicator qubit initialized to $\ket{0}$, with the ancilla qubits initialized to $\ket{\mathbf{0}}=\ket{00\ldots 0}$, and with $\ket{\psi}$ a valid witness (for $x\in L$), then the indicator qubit will remain $\ket{0}$ and nothing will happen; if, on the other hand, the witness/ancillae failed any of the verifiers, thus depolarizing the indicator qubit to be $\frac{1}{2}\id = \frac{1}{2}\selfketbra{0}+\frac{1}{2}\selfketbra{1}$, then $\finalexpander$ will act on the top registers, resulting in a highly mixed output (across all three registers). 
\end{enumerate}

Note that because $\finalexpander$ is an explicit $D_{\finalexpander}$-regular expander (where $D_{\finalexpander}$ is a constant), $\Phi$, being the composition of two explicit 8-regular superoperators and $\Lambda \finalexpander$, is manifestly explicit and $64D_{\finalexpander}$-regular (i.e. of constant degree).
We now proceed to show that $\Phi$ is indeed a good expander if $x\notin L$ (the NO case) but not if $x\in L$ (the YES case).


\subsection{Analysis of NO case}

First, consider the case in which $x\notin L$. We wish to show that $\Phi$ is a good expander, and therefore by Eq.~(\ref{eq:expander_defn}), that it sufficiently decreases the Frobenius norm of any input traceless matrix. As discussed earlier, we may therefore take the input state to be $\initstate$ for some matrices $A_i$ with $\tr{A_0}=0$, where $\sigma_i$ are the Pauli matrices on the indicator qubit register.

Both the witness and ancilla verifiers are controlled depolarizers, and we can analyse each of them in the same way using projection operators that act on some subspace of the system; specifically, we will use $Q = \sum_{\phi \text{ passes}}\selfketbra{\phi}$ that projects onto the states that pass the verifier and $P = \sum_{\phi \text{ fails}}\selfketbra{\phi}$ that projects onto the states that fail it.
For the ancilla verifier, these are $Q_a = \selfketbra{00\ldots0}_{anc}$ (more properly written as 
$Q_a = \mathbb{1}_{\text{wit}} \otimes \selfketbra{00\ldots0}_{anc} \otimes \mathbb{1}_{ind}$)
and
$P_a = \mathbb{1} - Q_a = \sum_{x\neq 00\ldots0} \selfketbra{x}_{anc}$.
For the witness verifier, $Q_w = V^\dag \selfketbra{1}_{top} V$ and $P_w = V^\dag \selfketbra{0}_{top} V$ (so that $P_w+Q_w=\id$). Here the subscript $top$ is used to indicate the top qubit register output from $V$.

Applying Eq.~(\ref{eq:controlled_depolarizer}) and linearity, the effect of a verifier unit on the input state $\sum_{i=0}^3 A_i \otimes \sigma_i$ is therefore
\begin{eqnarray*}
\label{eq:single_verifier_expanded}
F\left(\sum_{i=0}^3 A_i \otimes \sigma_i \right)
& = & 
\sum_{i=0}^3 \left[P A_i P \otimes \mathbb{1}\frac{\tr{\sigma_i}}{2} + Q A_i Q \otimes \sigma_i \right] \\
& = & 
P A_0 P \otimes \mathbb{1} + \sum_{i=0}^3 Q A_i Q \otimes \sigma_i .
\end{eqnarray*}

By linearity, it is easy to see that the effect of two such verifier units -- the ancilla verifier with projectors $\lbrace P_a,Q_a\rbrace$ and witness verifier with projectors $\lbrace P_w,Q_w\rbrace$ -- is
\begin{eqnarray*}
\label{eq:double_verifier}
&& F_w \circ F_a \left( \sum_{i=0}^3 A_i \otimes \sigma_i \right)\\
& = & 
  F_w \left( P_a A_0 P_a \otimes \mathbb{1} \right) 
+ F_w \left( \sum_{i=0}^3 Q_a A_i Q_a \otimes \sigma_i \right) \\
& = & 
\left(P_w P_a A_0 P_a P_w + Q_w P_a A_0 P_a Q_w + P_w Q_a A_0 Q_a P_w\right) \otimes \mathbb{1} + \sum_{i=0}^3 Q_w Q_a A_i Q_a Q_w \otimes \sigma_i \\
& = & 
\sum_P P A_0 P^\dag \otimes \mathbb{1} + \sum_{i=1}^3 Q A_i Q^\dag \otimes \sigma_i ,
\end{eqnarray*}
where the first sum is over $P\in\lbrace{P_wP_a,P_wQ_a,Q_wP_a,Q_wQ_a}\rbrace$ and where $Q$ is the single product $Q=Q_wQ_a$ and $Q^\dag=Q_aQ_w$.
Notice that the $i=0$ term (involving $\sigma_0 = \mathbb{1}$) in the second sum has been transferred to the first sum, thereby allowing the first sum to include all possible projection combinations.

We can rewrite this as 
\begin{equation}
\label{eq:double_verifier_simplified}
F_w \circ F_a \left( \sum_{i=0}^3 A_i \otimes \sigma_i \right) = C(A_0) \otimes \mathbb{1} + \sum_{i=1}^3 Q A_i Q^\dag \otimes \sigma_i
\end{equation}
where
\begin{equation*}
C(A_0) = \sum_P PA_0P^\dag = \sum_{R_w=P_w,Q_w}R_w \left( \sum_{R_a=P_a,Q_a} R_aA_0R_a \right) R_w = (G_w \circ G_a)(A_0)
\end{equation*}
is the composition of the pinching operators $G_j(B) = P_j B P_j + Q_j B Q_j$ applied to $A_0$. 

Since $C$ is the composition of pinching operators, Eqs.~(\ref{fact:pinch_trace}) and~(\ref{fact:pinch_norm}), along with Eq.~(\ref{fact:sum_greater_than_part}), tell us
\begin{equation}
\label{eq:trace_of_C}
\tr{C(A_0)} = \tr{A_0} = 0
\end{equation}
and
\begin{equation}
\label{eq:norm_of_C}
\frob{C(A_0)} \leqslant \frob{A_0} \leqslant \invsqrt2 \frob{ \sum_i A_i \otimes \sigma_i }.
\end{equation}

We are now ready to apply the final controlled expander, which by Eq.~(\ref{eq:controlled_expander_short}), with $P=\selfketbra{1}$ and $Q=\selfketbra{0}$, has the effect
\begin{equation*}
\Lambda\finalexpander\left(B \otimes b \right) 
= \finalexpander(B) \otimes \selfketbra{1}b\selfketbra{1} + B \otimes \selfketbra{0}b\selfketbra{0}.
\end{equation*} 
Applying this to the state Eq.~(\ref{eq:double_verifier_simplified}) we conclude that the effect of the map in Fig.~\ref{fig:whole_circuit} on the initial traceless matrix $\initstate$ is
\begin{equation*}
\label{eq:net_effect}
\Phi\left(\initstate\right) = C(A_0)\otimes\selfketbra{0} + \finalexpander\left(C(A_0)\right)\otimes\selfketbra{1} + QA_3Q^\dag\otimes\selfketbra{0} - \finalexpander(QA_3Q^\dag)\otimes\selfketbra{1}.
\end{equation*}

To show that $\Phi$ is a good quantum expander, we must show that it sufficiently decreases the Frobenius norm of its traceless input.
Since $\finalexpander$ is a $\kappa_{\finalexpander}$-contractive expander and $C(A_0)$ is traceless [see Eq.~(\ref{eq:trace_of_C})] we are guaranteed that
\begin{equation}
\label{eq:goodexpander}
\frob{\finalexpander\left(C(A_0)\right)} \leqslant \kappa_{\finalexpander} \frob{C(A_0)}.
\end{equation}
Applying the triangle inequality and Eqs.~(\ref{eq:goodexpander}),~(\ref{fact:expander_decreases}), and~(\ref{eq:norm_of_C}), we therefore have
\begin{eqnarray}
\label{eq:no_bound_part1}
\frob{\Phi\left(\initstate\right)}
& \leqslant &
\frob{C(A_0)} + \frob{\finalexpander\left(C(A_0)\right)} + \frob{QA_3Q^\dag} + \frob{\finalexpander(QA_3Q^\dag)}\notag\\
& \leqslant &
(1+\kappa_{\finalexpander})\frob{C(A_0)}  + 2\frob{QA_3Q^\dag}\notag\\
& \leqslant &
\frac{1+\kappa_{\finalexpander}}{\sqrt2}\frob{\initstate}  + 2\frob{QA_3Q^\dag}.
\end{eqnarray}
Note that we cannot make a claim similar to Eq.~(\ref{eq:goodexpander}) for $\finalexpander(QA_3Q^\dag)$ because $QA_3Q^\dag$ need not be traceless.

In QMA$(1,0)$ we are guaranteed that provided the ancillae are
initialized to be all 0's, no witness can pass the verifier (for a NO
instance). Mathematically, this guarantee is equivalent to saying that
$Q\equiv 0$. Consequently, the $QA_3Q^\dag$ vanishes and we are
done. In QMA$(a,b)$, however, we must upper bound $\frob{QA_3Q^\dag}$,
which we now proceed to do.

Because $x\notin L \in \text{QMA}(a,b)$ we are assured that for any purported witness $\ket{\psi}$, 
\begin{equation}
\label{eq:QMA_norm_b}
\left\|Q_w\ket{\psi}\ket{\textbf{0}}\right\|\leqslant \sqrt{b} .
\end{equation}

Because $Q_a$ projects onto the $\ketbra{\textbf{0}}{\textbf{0}}$ ancilla subspace, we may write
\begin{equation*}
Q_aA_3Q_a = \sum_{\psi_1,\psi_2}c(\psi_1,\psi_2)\ketbra{\psi_1}{\psi_2} \otimes \ketbra{\textbf{0}}{\textbf{0}}
\end{equation*}
where $\lbrace \ket{\psi_i} \rbrace$ is any orthonormal basis of the witness subspace. Note that because the witness register consists of $n_w$ qubits, $c(\psi_1,\psi_2)$ can be regarded as a matrix with dimension $N=2^{n_w}\times 2^{n_w}$. Thus using the triangle inequality and Eqs.~(\ref{fact:norms}) and~(\ref{eq:QMA_norm_b}),
\begin{eqnarray*}
\frob{QA_3Q^\dag}
& = &
\frob{\sum_{\psi_1,\psi_2}c(\psi_1,\psi_2) Q_w\ket{\psi_1}\ket{\textbf{0}} \bra{\psi_2}\bra{\textbf{0}}Q_w}\\
& \leqslant &
\sum_{\psi_1,\psi_2}\left|c(\psi_1,\psi_2)\right| \frob{\Big. Q_w\ket{\psi_1}\ket{\textbf{0}} \bra{\psi_2}\bra{\textbf{0}}Q_w}\\
& = &
\sum_{\psi_1,\psi_2}\left|c(\psi_1,\psi_2)\right| \frob{\Big. Q_w\ket{\psi_1}\ket{\textbf{0}}} \frob{\Big. Q_w \ket{\psi_2}\ket{\textbf{0}}}\\
& \leqslant &
\sum_{\psi_1,\psi_2}\left|c(\psi_1,\psi_2)\right| b .
\end{eqnarray*}
The matrix $c$ has $(2^{n_w})^2$ elements, so its 1-norm and 2-norm are related by
\begin{equation*}
\sum_{\psi_1,\psi_2}\left|c(\psi_1,\psi_2)\right| \leqslant 2^{n_w} \sqrt{\sum_{\psi_1,\psi_2}\left|c(\psi_1,\psi_2)\right|^2}
= 2^{n_w} \frob{Q_aA_3Q_a^\dag}.
\end{equation*}
But by Eqs.~(\ref{fact:projector_decreases}) and~(\ref{fact:sum_greater_than_part}),  $\frob{Q_aA_3Q_a^\dag} \leqslant \frob{A_3} \leqslant \invsqrt2\frob{\initstate}$; thus we conclude,
\begin{equation}
\label{eq:no_bound_part2}
\frob{QA_3Q^\dag} \leqslant \frac{2^{n_w}}{\sqrt2}\frob{\initstate} b.
\end{equation}
Although $2^{n_w}$ is exponential in ${n_w}$, recall that $b$ was chosen so that $2^{n_w+1}b\leqslant 0.1$. 
We conclude from Eqs.~(\ref{eq:no_bound_part1}) and~(\ref{eq:no_bound_part2}) that $\Phi$ is a $\beta$-contractive expander,
\begin{equation}
\label{eq:no_bound_final}
\frob{\Phi\left(\initstate\right)}
\leqslant
\beta\frob{\initstate},
\end{equation}
with
\begin{equation}
\label{eq:beta}
\beta = \frac{1+\kappa_{\finalexpander}+2^{n_w+1}b}{\sqrt2} < 0.85. 
\end{equation}
%


\subsection{Analysis of YES case}

Now consider the case in which $x\in L$. Since $L\in\text{QMA}(a,b)$ there exists a valid witness $\ket{\psi}$ such that
\begin{equation}
\label{eq:QMA_norm_a}
\left\|Q_w\ket{\psi}\ket{\textbf{0}}\right\|^2 \geqslant a.
\end{equation}
From this witness we construct the density matrix $\Psi = \selfketbra{\psi}\otimes\selfketbra{\mathbf{0}}\otimes\selfketbra{0}$. Because $\Psi$ passes the ancilla verifier unchanged and the witness verifier with very little change, $\Psi$ is almost a fixed point of our expander $\Phi$ (and indeed, for QMA$(1,0)$ it is a fixed point); intuitively, therefore, $\Phi$ is a poor expander. The matrix $\tilde{I}=\frac{1}{2^{n_w+n_a+1}}\id$ is certainly a fixed point (for any unital map); therefore the traceless matrix
\begin{equation*}
\label{eq:yescase_matrix}
A = \Psi -\tilde{I} = \selfketbra{\psi}\otimes\selfketbra{\mathbf{0}}\otimes\selfketbra{0} - \frac{1}{2^{n_w+n_a+1}}\id
\end{equation*}
is also expected to change very little under $\Phi$. By showing this to be the case, we will show that $\Phi$ is not an $\alpha$-contractive expander for an $\alpha$ that is polynomially separated from the $\beta$ found in the NO case.

Using an analysis similar to the previous case, it is easy to see that the effect of our circuit on $\Psi$ is
\begin{eqnarray*}
\Psi
& = &
\selfketbra{\psi}\otimes\selfketbra{\mathbf{0}}\otimes\selfketbra{0}\notag\\
& \overset{\text{Ancilla verifier}}{\xrightarrow{\hspace*{2cm}}} &
\selfketbra{\psi}\otimes\selfketbra{\mathbf{0}}\otimes\selfketbra{0}\notag\\
& \overset{\text{Witness verifier}}{\xrightarrow{\hspace*{2cm}}} &
P_w \big(\selfketbra{\psi}\otimes\selfketbra{\mathbf{0}}\big)P_w \otimes \frac{\id}{2} + Q_w\big(\selfketbra{\psi}\otimes\selfketbra{\mathbf{0}}\big)Q_w \otimes \selfketbra{0}\notag\\
& \overset{\text{Controlled $\finalexpander$}}{\xrightarrow{\hspace*{2cm}}} &
\frac{1}{2}\finalexpander\Big[P_w \big(\selfketbra{\psi}\otimes\selfketbra{\mathbf{0}}\big)P_w\Big] \otimes \selfketbra{1} \notag\\
&&~~~+ \frac{1}{2}P_w \big(\selfketbra{\psi}\otimes\selfketbra{\mathbf{0}}\big)P_w \otimes \selfketbra{0} \\
&&~~~+ Q_w\big(\selfketbra{\psi}\otimes\selfketbra{\mathbf{0}}\big)Q_w \otimes \selfketbra{0}. \notag
\end{eqnarray*}
Note that the three final terms are mutually orthogonal because $|0\>\<0|1\>\<1|=0$ and $P_w Q_w = 0$. Consequently, we have
\begin{eqnarray}
\label{eq:yes_norm_phipsi}
\frob{\Phi(\Psi)}^2 
&=&
\frac{1}{4}\frob{\finalexpander\Big[P_w \big(\selfketbra{\psi}\otimes\selfketbra{\mathbf{0}}\big)P_w\Big]}^2 \notag\\
&&~~~+\frac{1}{4}\frob{P_w \big(\selfketbra{\psi}\otimes\selfketbra{\mathbf{0}}\big)P_w}^2 \notag\\
&&~~~+\frob{Q_w\big(\selfketbra{\psi}\otimes\selfketbra{\mathbf{0}}\big)Q_w}^2 \notag\\
&\geqslant&
\frob{Q_w\big(\selfketbra{\psi}\otimes\selfketbra{\mathbf{0}}\big)Q_w}^2 \notag\\
&=&
\left\|Q_w\ket{\psi}\ket{\mathbf{0}}\right\|^4 \notag\\
&\geqslant&
a^2
\end{eqnarray}
where we have used Eq.~(\ref{fact:norms}) and Eq.~(\ref{eq:QMA_norm_a}).

Now, because $\Psi$ is a pure state density matrix, $\frob{A}^2 = \frob{\Psi - \tilde{I}}^2 = \tr{\Psi^2} + \tr{\tilde{I}^2} - 2\tr{\Psi \tilde{I}}$, using Eq.~(\ref{fact:frob_defn}), so that
\begin{equation}
\label{eq:yes_norm_A}
\frob{A}^2 = 1 - \frac{1}{2^{n_w+n_a+1}}.
\end{equation}
Thus, using that $\Phi$ is linear and trace-preserving, that $\Phi(\tilde{I})=\tilde{I}$, and Eqs.~(\ref{eq:yes_norm_phipsi}) and~(\ref{eq:yes_norm_A}), we have
\begin{eqnarray*}
\frob{\Phi(A)}^2
&=& \frob{\Phi(\Psi) - \Phi(\tilde{I})}^2 \notag\\
&=&
\tr{\Phi(\Psi)^\dag\Phi(\Psi)} + \tr{\tilde{I}^2} - \tr{\Phi(\Psi) \tilde{I}} - \tr{\Phi(\Psi)^\dag \tilde{I}} \notag\\
&=&
\frob{\Phi(\Psi)}^2 + \tr{\tilde{I}^2} - 2\tr{\Psi \tilde{I}} \notag\\
&\geqslant&
a^2 - \frac{1}{2^{n_w+n_a+1}}  \notag\\
&=&
\frob{A}^2 - (1-a^2)\notag\\
&>&
\left[1-\frac{8}{5}(1-a^2)\right] \frob{A}^2
\end{eqnarray*}
where in the last inequality we have used from Eq.~(\ref{eq:yes_norm_A}) that for $n_w\geqslant 1$ we have $\frac{5}{8} < \frob{A}^2 \leqslant 1$. 
Thus we conclude that $\Phi$ is not an $\alpha$-contractive expander,
\begin{equation}
\label{eq:yes_bound}
\frob{\Phi(A)} > \alpha \frob{A},
\end{equation}
with 
\begin{equation}
\label{eq:alpha}
\alpha = \sqrt{1-\frac{8}{5}(1-a^2)} > 0.98 . 
\end{equation}
Note that $\alpha$ and $\beta$ are constants, and therefore certainly polynomially separated.

%
%

\section{Conclusion}
We have presented a new computational problem, \emph{quantum
  non-expander}, and proved that it is QMA-complete.  
This gives some insight into the computational complexity of 
estimating mixing rates of quantum channels and open quantum systems. 

In contrast to the plethora of natural NP-complete problems,
very few problems have been shown to be QMA-complete.  We hope that 
it may be possible to find new QMA-complete problems, 
using reductions from the quantum non-expander problem.

%
%

\subsection*{Acknowledgments}

This work was supported in part by the U.S. Department of Energy under cooperative research agreement Contract Number DE-FG02-05ER41360, as well as 
by the National Science Foundation Science and Technology Center for Science of Information under grant CCF-0939370.  
P.W. gratefully acknowledges the support from the NSF CAREER Award CCF-0746600.  

Contributions to this work by NIST, an agency of the US government, are not subject to copyright laws.

\appendix
\section{Appendix}

\subsection{Master equation for a quantum system coupled to a bath}
\label{apx-thermal}
In this section we derive the master equation (\ref{eqn-master}), given the system-bath Hamiltonian specified in (\ref{eqn-ham1}), (\ref{eqn-ham2}) and (\ref{eqn-ham3}).  We follow the arguments of sections 3.3 and 3.4 in \cite{BP-book}.  

First, define new operators $A_{\alpha\sigma}$ and $B_{\alpha\sigma}$ (for $\alpha = 1,\ldots,D$ and $\sigma = 0,1$):
\[
A_{\alpha\sigma} = \frac{1}{\sqrt{2}} (-i)^\sigma (U_\alpha + (-1)^\sigma U_\alpha^\dagger), \qquad
B_{\alpha\sigma} = \frac{1}{\sqrt{2}} i^\sigma (f_\alpha + (-1)^\sigma f_\alpha^\dagger).
\]
Then we can write the interaction Hamiltonian in the form 
\[
H_I = \sum_{\alpha\sigma} A_{\alpha\sigma} \otimes B_{\alpha\sigma}.
\]
This form is convenient because $A_{\alpha\sigma}$ and $B_{\alpha\sigma}$ are Hermitian.

In the weak-coupling limit ($\varepsilon \rightarrow 0$), one gets the following master equation (equation 3.140 in \cite{BP-book}, simplified using the fact that $H_S = 0$):
\begin{equation} \label{eqn-master2}
\frac{d}{dt} \rho_S(t) = -i[H_{LS}, \rho_S(t)] + \mathcal{D}(\rho_S(t)), 
\end{equation}
where $H_{LS}$ is the ``Lamb shift'' Hamiltonian and $\mathcal{D}$ is the dissipator, 
\[
H_{LS} = \sum_{\alpha\beta\sigma\tau} S_{\alpha\beta\sigma\tau} A_{\alpha\sigma}^\dagger A_{\beta\tau}, \qquad
\mathcal{D}(\rho_S) = \sum_{\alpha\beta\sigma\tau} \gamma_{\alpha\beta\sigma\tau}
 \Bigl( A_{\beta\tau} \rho_S A_{\alpha\sigma}^\dagger
 - \frac{1}{2} \lbrace A_{\alpha\sigma}^\dagger A_{\beta\tau}, \rho_S \rbrace \Bigr),
\]
and the coefficients $S_{\alpha\beta\sigma\tau}$ and $\gamma_{\alpha\beta\sigma\tau}$ are given by 
\[
S_{\alpha\beta\sigma\tau} = \frac{1}{2i} (\Gamma_{\alpha\beta\sigma\tau} - \Gamma_{\beta\alpha\tau\sigma}^*), \qquad
\gamma_{\alpha\beta\sigma\tau} = \Gamma_{\alpha\beta\sigma\tau} + \Gamma_{\beta\alpha\tau\sigma}^*, 
\]
where $\Gamma_{\alpha\beta\sigma\tau}$ are the one-sided Fourier transforms (evaluated at frequency 0) of the bath correlation functions, 
\[
\Gamma_{\alpha\beta\sigma\tau} = \int_0^\infty ds \langle B_{\alpha\sigma}^\dagger(s) B_{\beta\tau}(0) \rangle, \qquad
B_{\alpha\sigma}(t) = e^{iH_Bt} B_{\alpha\sigma} e^{-iH_Bt}.
\]

We can evaluate the bath correlation functions, using the fact that the bath is in a thermal state at temperature $T$.  After some algebra, we get 
\[
\begin{split}
\langle B_{\alpha\sigma}^\dagger(s) B_{\beta\tau}(0) \rangle
 &= \frac{1}{2} i^\sigma i^\tau \frac{1}{|\Omega|} \sum_{kk'}
 \Bigl( e^{-is\omega_k} \langle b_{\alpha k} b_{\beta k'} \rangle
 + e^{-is\omega_k} (-1)^\tau \langle b_{\alpha k} b_{\beta k'}^\dagger \rangle \\
 &\qquad\qquad\qquad\qquad + (-1)^\sigma e^{is\omega_k} \langle b_{\alpha k}^\dagger b_{\beta k'} \rangle
 + (-1)^\sigma e^{is\omega_k} (-1)^\tau \langle b_{\alpha k}^\dagger b_{\beta k'}^\dagger \rangle \Bigr) \\
 &= \frac{1}{2} i^\sigma i^\tau \delta_{\alpha\beta} \frac{1}{|\Omega|} \sum_k
 \Bigl( (-1)^\sigma e^{is\omega_k} N(\omega_k) + e^{-is\omega_k} (-1)^\tau (1+N(\omega_k)) \Bigr), 
\end{split}
\]
where $N(\omega_k) = \frac{1}{\exp(\omega_k/T)-1}$.  We take a continuum limit, replacing the sum $\frac{1}{|\Omega|} \sum_k$ by an integral $\int_\Omega dk$; this amounts to using a bath with infinitely many modes, and is necessary to obtain irreversible behavior of the system.  

We then substitute the above expression into the definition of $\Gamma_{\alpha\beta\sigma\tau}$:
\[
\Gamma_{\alpha\beta\sigma\tau}
 = \frac{1}{2} i^\sigma i^\tau \delta_{\alpha\beta} \int_0^\infty ds \int_\Omega dk \Bigl( (-1)^\sigma e^{is\omega_k} N(\omega_k) + e^{-is\omega_k} (-1)^\tau (1+N(\omega_k)) \Bigr).
\]
We can simplify the above formula by exchanging the integrals and using the identity 
$\int_0^\infty ds e^{-ixs} = \pi\delta(x) - i \cdot PV(\tfrac{1}{x})$, 
where $\delta(x)$ is the Dirac distribution and $PV(\tfrac{1}{x})$ is the Cauchy principal value (equation 3.202 in \cite{BP-book}).  We then get:
\[
\begin{split}
\Gamma_{\alpha\beta\sigma\tau}
 &= \frac{1}{2} i^\sigma i^\tau \delta_{\alpha\beta} \int_\Omega 
    \Bigl( (-1)^\sigma N(\omega_k) \int_0^\infty e^{is\omega_k} ds 
         + (-1)^\tau (1+N(\omega_k)) \int_0^\infty e^{-is\omega_k} ds \Bigr) dk \\
 &= \frac{1}{2} i^\sigma i^\tau \delta_{\alpha\beta}
    \Bigl( (-1)^\sigma \pi N(0) + (-1)^\sigma i \cdot PV \int_\Omega \frac{N(\omega_k)}{\omega_k} dk \\
 &\qquad\qquad\quad
         + (-1)^\tau \pi(1+N(0)) - (-1)^\tau i \cdot PV \int_\Omega \frac{1+N(\omega_k)}{\omega_k} dk \Bigr).
\end{split}
\]
In particular, $\Gamma_{\alpha\beta\sigma\tau}$ can be written in the form 
\[
\Gamma_{\alpha\beta\sigma\tau}
 = \frac{1}{2} i^\sigma i^\tau \delta_{\alpha\beta} \Bigl( (-1)^\sigma Q_0 + (-1)^\tau Q_1 \Bigr), 
\]
where the coefficients $Q_0$ and $Q_1$ are complex numbers with positive real part.

We can now calculate the ``Lamb shift'' Hamiltonian $H_{LS}$ as follows:
\[
S_{\alpha\beta\sigma\tau}
 = \frac{1}{2i} \cdot \frac{1}{2} i^\sigma i^\tau \delta_{\alpha\beta}
 \Bigl( (-1)^\sigma (Q_0-Q_0^*) + (-1)^\tau (Q_1-Q_1^*) \Bigr),
\]
\[
\begin{split}
H_{LS}
 &= \frac{Q_0-Q_0^*}{4i} \sum_\alpha \Bigl( \sum_\sigma i^\sigma (-1)^\sigma A_{\alpha\sigma}^\dagger \Bigr)
 \Bigl( \sum_\tau i^\tau A_{\alpha\tau} \Bigr) \\
 &+  \frac{Q_1-Q_1^*}{4i} \sum_\alpha \Bigl( \sum_\sigma i^\sigma A_{\alpha\sigma}^\dagger \Bigr)
 \Bigl( \sum_\tau i^\tau (-1)^\tau A_{\alpha\tau} \Bigr) \\
 &= \frac{Q_0-Q_0^*}{4i} \sum_\alpha \sqrt{2} U_\alpha^\dagger U_\alpha \sqrt{2}
 + \frac{Q_1-Q_1^*}{4i} \sum_\alpha \sqrt{2} U_\alpha U_\alpha^\dagger \sqrt{2} \\
 &= \frac{Q_0-Q_0^*}{2i} D I + \frac{Q_1-Q_1^*}{2i} D I.
\end{split}
\]
So $H_{LS}$ is a multiple of the identity, and it contributes nothing when we substitute it into the master equation (\ref{eqn-master2}).

Finally we can calculate the dissipator $\mathcal{D}$.  First, 
\[
\gamma_{\alpha\beta\sigma\tau}
 = \frac{1}{2} i^\sigma i^\tau \delta_{\alpha\beta}
 \Bigl( (-1)^\sigma (Q_0+Q_0^*) + (-1)^\tau (Q_1+Q_1^*) \Bigr).
\]
We substitute this into the definition of $\mathcal{D}$, and simplify it in the same way as we did for $H_{LS}$.  This yields 
\[
\mathcal{D}(\rho_S)
 = (Q_0+Q_0^*) \sum_\alpha \Bigl( U_\alpha \rho_S U_\alpha^\dagger - \rho_S \Bigr)
 + (Q_1+Q_1^*) \sum_\alpha \Bigl( U_\alpha^\dagger \rho_S U_\alpha - \rho_S \Bigr).
\]
Note that $Q_0+Q_0^*$ and $Q_1+Q_1^*$ are positive real numbers.  We substitute this into the master equation (\ref{eqn-master2}).  This completes our proof of (\ref{eqn-master}).

\subsection{Controlled expanders}
\label{sec:appendix}
In this appendix, we outline how we obtain the requisite controlled expander $\Lambda\finalexpander$ needed for section~\ref{sec:QMAhard}. We use the results of Ben-Aroya, Schwartz, and Ta-Shma~\cite{BST10}, whose Theorem 4.3 and 4.6 give the following result.

%
%
%

\begin{theorem}
There exists an integer $D_0$ such that for every $D>D_0$ and for every integer $t>0$,
there exists a explicit $\lambda_t$-contractive expander of degree $D^2$ on a space of dimension $D^{8t}$
where
$\lambda_t \leqslant \lambda + c\lambda^2$ with $c$ a constant and $\lambda=\frac{4\sqrt{D-1}}{D}$.
\end{theorem}

We will additionally use the following result, which follows directly from the definition. Here we use the notation that $\mathcal{F}^r$ denotes the $r$-fold composition of $\mathcal{F}$.
\begin{proposition}
If $\mathcal{F}$ is a $\lambda$-contractive expander of degree $D$ on a space of size $N$, then for any positive integer $r$, $\mathcal{F}^r$ is a $\lambda^r$-contractive expander of degree $D^r$ on a space of size $N$.
\end{proposition}

In section~\ref{sec:QMAhard} we require an $\kappa_{\finalexpander}$-contractive expander $\finalexpander$ with $\kappa_{\finalexpander}\leqslant 0.1$ on a space of size $N=2^{n_w+n_a}$. Note that $N$ is actually allowed to exceed $2^{n_w+n_a}$ since we can always have extra input ancillae that do nothing but are acted upon by the final controlled expander $\Lambda\finalexpander$.

Fix $D$ to be any power of 2 larger than $D_0$. Then $\lambda=\frac{4\sqrt{D-1}}{D}<1$ is fixed. Let $r$ be such that 
$(\lambda+c\lambda^2)^r \leqslant 0.1$.
Let $t = \ceil{\frac{n_w+n_a}{8\log_2{D}}} = \frac{n_w+n_a+n_{extra}}{8\log_2{D}}$ for some $n_{extra}<8\log_2{D}$. 

Using the above theorem we are guaranteed the existance of a $\lambda_t^r$-contractive expander of degree $D^{2r}$ on a space of size $D^{8t} = 2^{n_w+n_a+n_{extra}}$, where
$D$ and $r$ are constants and $\lambda_t^r\leqslant 0.1$.

%
%

\end{document}